\documentclass[a4paper,11pt]{article}
\usepackage{pos}
\usepackage[utf8]{inputenc}
\usepackage{graphicx}
\usepackage{tabularx}
\usepackage{hyphenat}
\usepackage{amsmath,amsfonts,amsthm,latexsym} 
\usepackage{slashed}
\usepackage{siunitx}
\usepackage{caption}
\usepackage{subcaption}
\usepackage{lineno}
\usepackage{hyperref}
\usepackage{mathtools}
\usepackage{placeins} 

\title{Quenched Static force from generalized Wilson loops with gradient flow}

\author*[a,b]{Julian Mayer-Steudte}

\affiliation[a]{Physik Department, Technische Universit\"at M\"unchen,
James-Franck-Strasse 1, 85748 Garching, Germany}

\affiliation[b]{Munich Data Science Institute, Technische Universit\"at M\"unchen, \\
Walther-von-Dyck-Strasse 10, 85748 Garching, Germany}

\emailAdd{julian.mayer-steudte@tum.de}

\abstract{We compute the static force on the lattice in the quenched case directly through generalized Wilson loops. We modify the Wilson loop by inserting an $E$-field component on one of the temporal Wilson lines. However, chromo-field components prevent us from performing the continuum limit properly, hence, we use gradient flow to renormalize the field insertion. As a result, we obtain continuum results and compare them to perturbative expression to extract $\Lambda_0$, and we predict the value $\sqrt{8t_0} \Lambda_{\overline{\textrm{MS}}}^{n_f=0} =0.629^{+22}_{-26}$. This work serves as preparation for similar operators with field insertions required in nonrelativistic effective field theories.}

\FullConference{The 40th International Symposium on Lattice Field Theory (Lattice 2023)\\
July 31st - August 4th, 2023\\
Fermi National Accelerator Laboratory\\}

\begin{document}
\maketitle

\section{Introduction}
Precise theoretical knowledge about Standard model parameters is necessary to test the Standard model, and to determine the running of the strong coupling $\alpha_S$. One of the parameters is the QCD scale $\Lambda_\mathrm{QCD}$ which determines the $\alpha_S$-running. To extract $\Lambda_\mathrm{QCD}$, we compare perturbative results with non-perturbative results from the lattice, in a regime where both approaches have the same validity. Calculating the force between a static quark-antiquark is one possibility, which is traditionally carried out by computing the static force on the lattice from a numerical derivative of the static energy. In this work, we measure the static force directly through a generalized Wilson loop with a chromoelectric field $E$ insertion, avoiding the step of performing a numerical derivative, where additional systematic uncertainties arise. Similar objects with field insertions are required for correlators needed in nonrelativistic effective field theory calculations. However, the discretized lattice $E$-field insertion includes a non-trivial lattice spacing behavior from gluonic self-interactions, slowing the convergence to the continuum limit. This can be solved by a renormalization of the $E$-field insertion. We use gradient flow for both, to renormalize and to improve the signal-to-noise ratio of our observables, and thereafter, to perform the continuum limit. In the continuum limit, we use different methods to extract the zero flow time limit. This work was recently published in a preprint paper \cite{Brambilla:2023fsi} and we summarize some key findings in this proceeding.

\section{Theoretical Background} \label{sec:theoretical_background}

The static force is obtained through the static energy $V(r)$ in Euclidean QCD, which is related to a Wilson loop $W_{r\times T}$ with spatial extent $r$ and temporal extent $T$ as
\begin{linenomath}\begin{align}
    V(r) &= -\lim_{T\rightarrow \infty} \frac{\ln\langle \mathrm{Tr}(W_{r\times T})\rangle}{T} = - \frac{1}{a}\lim_{T\rightarrow \infty} \frac{\langle \mathrm{Tr}(W_{r\times(T+a)})\rangle}{\langle \mathrm{Tr}(W_{r\times T})\rangle}\,,\\
    W_{r\times T} &= P\left\{ \exp\left( i\oint _{r\times T} dz_\mu gA_\mu \right) \right\} = \underset{{(i,\mu)\in\mathcal{C}}}{\prod} U_{\mu}(n_i),\label{eq:wilson_loop_definition}
\end{align}\end{linenomath}
where $a$ is the lattice spacing, $g$ the strong coupling constant, $A_\mu$ the gluon fields, and $P$ the path ordering operator. The product over $(i,\mu)\in\mathcal{C}$ produces the path ordered product of link variables $U_\mu(n_i)$ along the closed loop $\mathcal{C}$ representing the Wilson loop on the lattice. The trace operation $\mathrm{Tr}(...)$ is the normalized color trace. A derivative of the static energy defines the static force:
\begin{linenomath}\begin{align}
    F_{\partial V}(r) = \frac{\partial V}{\partial r} = \lim_{a\rightarrow 0} \frac{V(r+a)-V(r-a)}{2a},
\end{align}\end{linenomath}
where the term in the $\lim$-part on the right-hand side corresponds to the symmetric numerical derivative performed on the lattice. Other methods, including derivatives of interpolating functions, are possible, but infer additional systematic uncertainties driven by the interpolation method.

The static force in perturbation theory is known up to next-to-next-to-next-to leading logarithmic order (N$^3$LL) ~\cite{Brambilla:1999qa,Pineda:2000gza,Brambilla:2006wp,Anzai:2009tm,Smirnov:2009fh}. With renormalized coupling in $\overline{\textrm{MS}}$ scheme, the QCD scale and the strong coupling $\alpha_S$ are defined in $\overline{\textrm{MS}}$. Hence, in the quenched case, the $\alpha_S$-running is parametrized by $\Lambda_0\equiv\Lambda_{\overline{\textrm{MS}}}^{n_f=0}$. Since $r$ is the only scale within a static quark-antiquark pair, it is a natural choice to set the scale as $\mu = 1/r$.

Instead of performing the numerical derivative of the static energy, we measure the force directly ~\cite{Vairo:2015vgb,Vairo:2016pxb,Brambilla:2000gk}:
\begin{linenomath}\begin{align}
    F_E(r) &= -\lim _{T\rightarrow\infty}\frac{i}{\langle \mathrm{Tr}(W_{r\times T} \rangle}\left\langle\mathrm{Tr}\left( P\left\{ \exp\left( i\oint _{r\times T} dz_\mu gA_\mu \right) \mathbf{\hat{r}}\cdot g\mathbf{E}(\mathbf{r}, t^*) \right\} \right) \right\rangle \\
    &= -\lim _{T\rightarrow\infty} i \frac{\langle \mathrm{Tr}\{ PW_{r\times T}gE_j(\mathbf{r},t^* \} \rangle}{\langle \mathrm{Tr}(W_{r\times T} \rangle} \label{eq:static_force_E_insertion_definition}.
\end{align}\end{linenomath}
The new object is the chromoelectric field component $\mathbf{E}$, inserted in one of the temporal Wilson lines of the Wilson loop at temporal slice $t^*$. The force does not depend on the insertion location $t^*$, however, to reduce the interaction of the insertion with the spatial Wilson lines in our lattice correlator, we choose $t^*$ to be the middle of the temporal Wilson line. Finally, we take only that component from $\mathbf{E}$ which is the same as for the separation axis of the quark-antiquark pair by multiplying it with the normalized direction vector $\mathbf{\hat{r}}$.

We discretize the chromo field components on the lattice with a clover discretization as 
\begin{linenomath}\begin{align}
    a^2F_{\mu\nu}&=\frac{-i}{8}(Q_{\mu\nu}-Q_{\nu\mu}) \label{eq:clover_leaf_definition}\\
    Q_{\mu\nu} &= U_{\mu,\nu}+U_{\nu,-\mu}+U_{-\mu,-\nu}+U_{-\nu,\mu} = Q_{\nu\mu}^\dagger
\end{align}\end{linenomath}
where $U_{\mu\nu}$ is a plaquette in the $\mu$-$\nu$-plane. In addition, we manually make the field components traceless, corresponding to an $a^2$-improvement \cite{Bilson-Thompson:2002xlt}. The electric field components are given through $a^2E_i=-a^2F_{i,4}$. At tree level, this discretization corresponds to the symmetric finite difference derivative defined above. 

However, the $E$-field discretization induces a non-trivial and slow convergence to the continuum limit, seen in a lattice perturbation calculation \cite{Lepage:1992xa}. This issue is absent in the force obtained through the derivative of the static energy. In this way, we can set a renormalization condition
\begin{linenomath}\begin{align}
    Z_E F_E(r) = F_{\partial V}(r).\label{eq:E_field_renormalization_condition}
\end{align}\end{linenomath}
where the non-trivial behavior is absorbed into $Z_E$. If $Z_E=1$, we may assume that $F_E$ behaves trivially in the continuum limit. $Z_E$ was non-perturbatively studied in \cite{Brambilla:2021wqs}, and it was found that it has only a low $r$-dependence.

We use gradient flow ~\cite{Narayanan:2006rf, Luscher:2009eq, Luscher:2010iy}, which introduces an additional scale, the flow radius $\sqrt{8\tau_F}$, and also an additional reference scale, $t_0$, to renormalize the $E$-field insertion, to perform the continuum limit, and to compare to perturbative results for extracting $\alpha_S$. The force in continuum perturbation theory at finite flow time is known up to 1-loop order \cite{Brambilla:2021egm}. The full expression can be expanded up to leading order in $\tau_F/r^2$:
\begin{linenomath}\begin{align}
r^2F(r,\tau_F) \approx r^2F(r,\tau_F=0) + \frac{\alpha_s^2C_F}{4\pi}\left[ -12\beta_0 - 6C_A c_L\right] \frac{\tau_F}{r^2}
\label{eq:force_tauf_r_expanded}
\end{align}\end{linenomath}
with $c_L=-22/3$, $C_F=(N_C^2-1)/(2N_C)$, $C_A=N_C$, $N_C=3$ the number of colors, $\beta_0=\frac{11}{3}C_A-\frac{2}{3}n_f$, and $n_f$ the number of flavors. We remark here that $[-12\beta_0 - 6C_A c_L]=8n_f$, which is 0 in this study ($n_f=0$); hence, the static force is constant at small flow time. $r^2F(r,\tau_F=0)$ is the full 1-loop expression of the force at zero flow time times $r^2$.

With the gradient flow scale, we have the choice between both scales, $r$ and $\sqrt{8\tau_F}$. From a perturbative argument, it is suggested to take an average of both scales as $\mu=1/\sqrt{r^2+8\tau_F}$. However, in this study, we define a parametrized mixture of both scales as
\begin{linenomath}
    \begin{align}
        \mu = \frac{1}{\sqrt{r^2 + 8b\tau_F}} \label{eq:mu_scaling_at_finite_tf}
    \end{align}
\end{linenomath}
where the parameter $b$ fixes the weight of the gradient flow scale included in $\mu$. In the zero flow time limit $\tau_F\rightarrow 0$, the scale approaches $\mu\rightarrow 1/r$ for any $b$, which is the natural choice at zero flow time.

In this proceeding, we parametrize the perturbative expression of the force with $\Lambda_0$ in two cases, for 1-loop at finite and zero flow time, and for 3-loop with leading ultrasoft resummation. We label the former case as F1l, and the latter one as F3lLus. The higher order perturbation regime is crucial for a reliable $\Lambda_0$-extraction. To benefit from the high-order knowledge of the force even at finite flow time, we model the force at zero flow time with F3lLus, and the finite flow time effects with the finite flow time expression of F1l, demanding that for $\tau_F\rightarrow 0$ that it converges to F3lLus. For a comparison among different orders, see the full study \cite{Brambilla:2023fsi}.

\section{Results}\label{sec:results}
\begin{table}
    \centering
    \begin{tabular}{c|c|c|c|c|c|c}
        $N_S$ & $N_T$ & $\beta$ & $a$ [fm] & $t_0/a^2$ &  $N_\mathrm{conf}$ & Label \\\hline
        $20^3$ & $40$ & 6.284 & 0.060 & 7.868(8) & 6000 &L20\\\hline
        $26^3$ & $56$ & 6.481 & 0.046 & 13.62(3) & 6000 & L26\\\hline
        $30^3$ & $60$ & 6.594 & 0.040 & 18.10(5) & 6000 & L30\\\hline
        $40^3$ & $80$ & 6.816 & 0.030 & 32.45(7) & 3300 & L40\\\hline
    \end{tabular}
    \caption{The parameters for the lattice ensembles. The $t_0$ scale was extracted with a smaller subset of the lattice configurations. The lattice spacing in physical units is determined by the scale setting from \cite{Necco:2001xg} with $r_0=\SI{0.5}{fm}$.}
    \label{tab:lattice_parameters}
\end{table}

The lattice computations were carried out in pure gauge with overrelaxation and heatbath algorithm to generate the ensembles, and a fixed step size \cite{Luscher:2010iy}, or an adaptive step size algorithm \cite{Fritzsch:2013je, Bazavov:2021pik} to solve the gradient flow equations numerically. The reference scale $t_0$ is obtained in lattice units from the action density with clover discretization Eq. \eqref{eq:clover_leaf_definition}. We use this reference scale to perform the continuum limit as $a^2/t_0\rightarrow 0$. The full lattice simulation parameters and the determined reference scales are shown in Table \ref{tab:lattice_parameters}.

\begin{figure}
\centering
\begin{subfigure}{.47\textwidth}
    \centering
    \includegraphics[width=\textwidth]{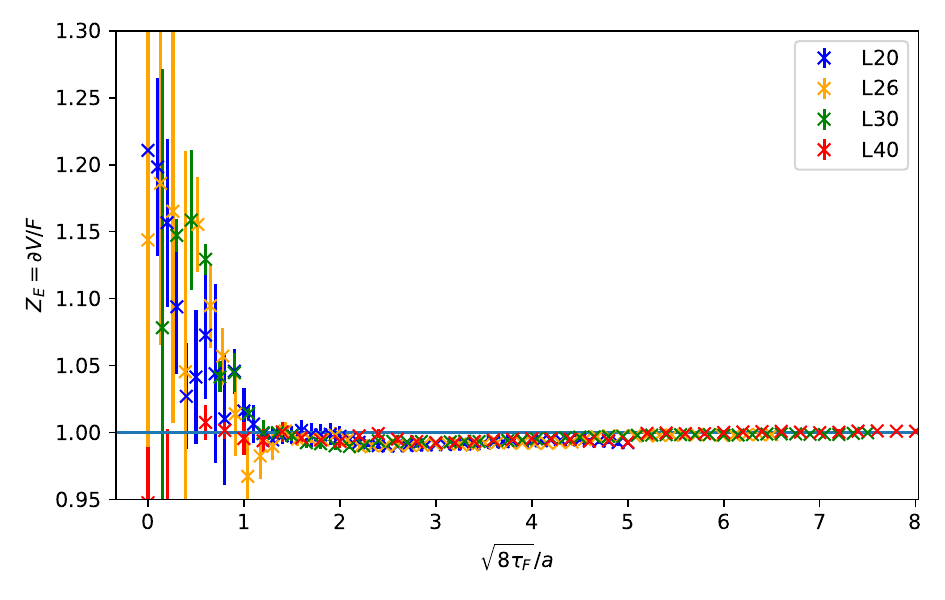}
\end{subfigure}
\begin{subfigure}{.47\textwidth}
    \centering
    \includegraphics[width=\textwidth]{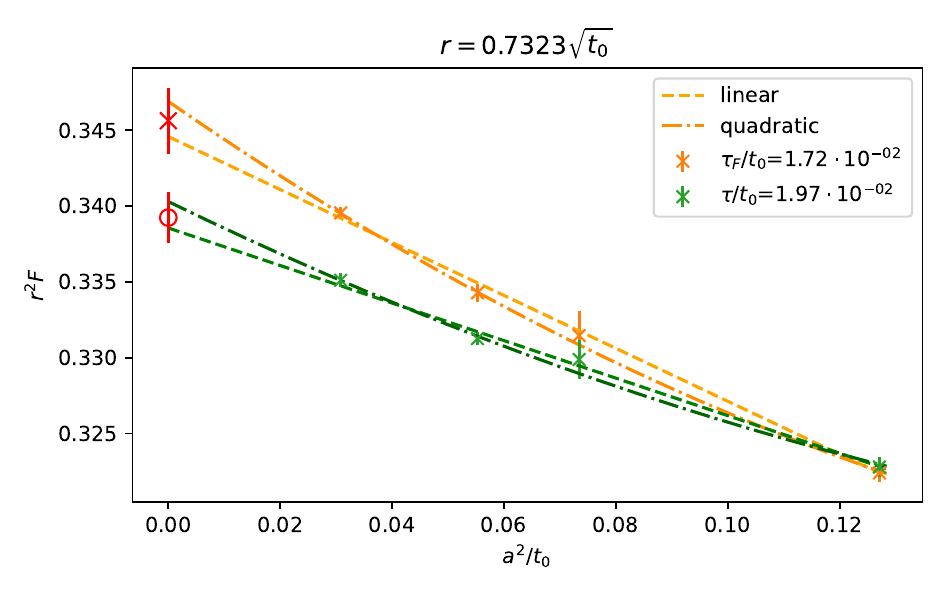}
\end{subfigure}
\caption{Left: The non-perturbative determination of $Z_E$ for all lattice sizes as a flow-time-dependent function. Right: An example of the weighted continuum limit at two different flow times at fixed $r$.}
\label{fig:ZE_and_cont_limit_example}
\end{figure}

The $T\rightarrow\infty$ limit is performed as plateau fit with an Akaike information criterion based procedure \cite{Jay:2020jkz}, the full procedure is explained in detail in the main publication \cite{Brambilla:2023fsi}.

To show the renormalizing property of the gradient flow, we determine $Z_E$ non-perturbatively by solving Eq. \eqref{eq:E_field_renormalization_condition} for $Z_E$ and represent it as a function of flow time. The left side of Fig. \ref{fig:ZE_and_cont_limit_example} shows $Z_E(\tau_F)$ for the different lattice sizes. We identify $Z_E=1$, which is required to perform a reliable continuum limit, within a \SI{1}{\percent} range for flow radii larger than one lattice spacing, i.e. $\sqrt{8\tau_F}>a$, which defines a minimum amount of required flow time. The discrepancy for $Z_E$ from 1 originates in the systematic uncertainty of the numerical derivative of the static energy, which can be seen by comparing it to other derivative methods. The right side of Fig. \ref{fig:ZE_and_cont_limit_example} shows an example of the continuum limit in the valid regime where $Z_E\approx 1$ and $\chi^2/dof$ gives a trustable result. We use Akaike weighted averages \cite{Jay:2020jkz} of continuum limits in a linear and quadratic in $a^2/t_0$ fit, and we restrict to continuum results with at least either the linear or quadratic fit satisfy $\chi^2/dof<3$. Since the minimal amount of required flow time is finite, we still need to extract the zero flow time physics from the finite flow time continuum results.

\begin{figure}
\centering
\begin{subfigure}{.47\textwidth}
    \centering
    \includegraphics[width=\textwidth]{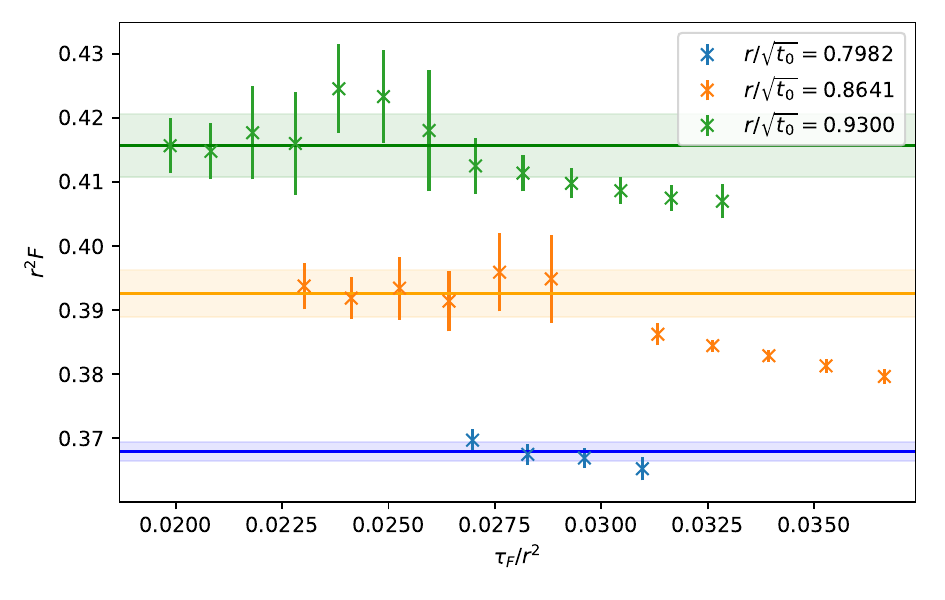}
\end{subfigure}
\begin{subfigure}{.47\textwidth}
    \centering
    \includegraphics[width=\textwidth]{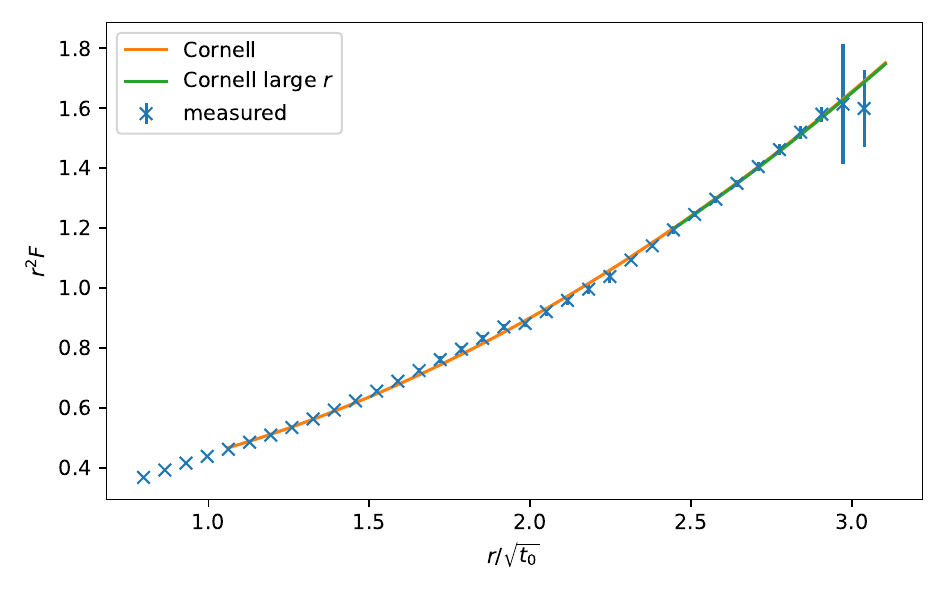}
\end{subfigure}
\caption{Left: An example of the constant zero flow time limit of the force at smallest possible $r$. \\Right: The static force at larger $r$ with Cornell fits.}
\label{fig:force_constant_zero_flow_time}
\end{figure}

We can choose the order by taking the $\tau_F\rightarrow0$ extrapolation of the force first or taking directly the $\Lambda_0$-fit to the force at finite flow time. The first approach, we present here, takes advantage of the fact that Eq. \eqref{eq:force_tauf_r_expanded} has a constant behavior of the force at small flow times. We identify flow time ranges where $F(\tau_F)$ stays constant within errors, and perform constant zero flow time limits. The left side of Fig. \ref{fig:force_constant_zero_flow_time} shows an example of a constant zero flow time limit at the smallest possible $r$. This method also works for larger $r$, and we can perform a Cornell fit to the extrapolated data, shown on the right side of Fig. \ref{fig:force_constant_zero_flow_time}. If we restrict the Cornell fit to $r/\sqrt{t_0}>2.3$, we obtain
\begin{linenomath}
    \begin{align}
        r^2F^\mathrm{Cornell}(r) &= A + \sigma r^2 \\
        A &= \num{0.268(33)} \\
        \sigma t_0 &= 0.154(6)
    \end{align}
\end{linenomath}
where $A$ is a dimensionless parameter, and $\sigma$ the string tension. At smallest $r$, we fit the perturbative expression of the force to the extrapolated lattice data. For F1l, we obtain $\sqrt{8t_0}\Lambda_0^\mathrm{F1l}=0.821(5)$, and for F3lLus $\sqrt{8t_0}\Lambda_0^{\mathrm{F3lLus}}=0.635(4)$.

\begin{figure}
\centering
\begin{subfigure}{.47\textwidth}
    \centering
    \includegraphics[width=\textwidth]{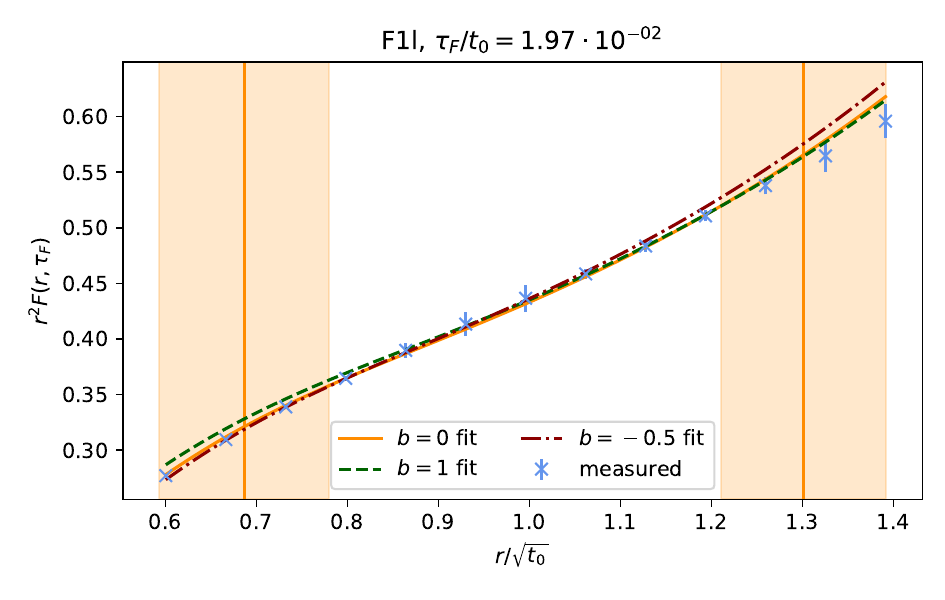}
\end{subfigure}
\begin{subfigure}{.47\textwidth}
    \centering
    \includegraphics[width=\textwidth]{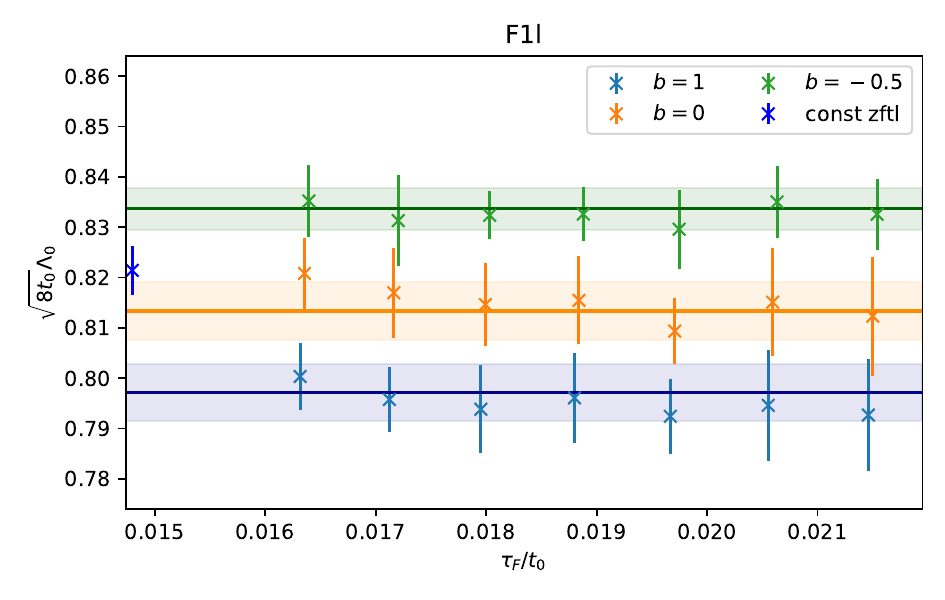}
\end{subfigure}
\caption{Left: Fit of the 1-loop perturbative expression to the lattice data. We present a scaling behavior with $b=0$ from Eq. \eqref{eq:mu_scaling_at_finite_tf}, and a fit where $b$ is treated as an additional fit parameter. The vertical lines with bands correspond to the average fit range bounds. Right: The extracted $\Lambda_0$ at different flow times for different $b$ with a constant zero flow time limit. The result of the previous approach is included as well.}
\label{fig:finite_fixed_tf_force_fit}
\end{figure}

For the second way to extract $\Lambda_0$ we fit the perturbative expressions at finite flow time. The perturbative expressions at finite flow time are parametrized by $\Lambda_0$, hence, fitting $\Lambda_0$ at finite $\tau_F$ already includes the zero flow time limit. We keep the flow time fixed and perform a fit along the $r$-axis. The left side of Fig. \ref{fig:finite_fixed_tf_force_fit} shows an example for F1l, where we chose the scale according to Eq. \eqref{eq:mu_scaling_at_finite_tf} for different values of $b$. We perform the fit within different $r$-ranges and use the weighted average with an Akaike information criterion \cite{Jay:2020jkz} for the fit result. The vertical lines with the bands represent the weighted average fit ranges and their deviations, respectively. The right side of Fig. \ref{fig:finite_fixed_tf_force_fit} shows the fit results for $\Lambda_0$ at different flow times. As expected, the $\tau_F$-dependence of $\Lambda$ is negligible within the errors, since every $\Lambda_0$ represents an independent zero flow time limit, and we perform a constant fit over all $\Lambda_0$, which is represented by the horizontal lines and bands. In addition, the plot includes the result of the previous approach, where we perform a constant zero flow time limit of the force first. We remark that the previous approach agrees with this one within the errors. To estimate the systematic uncertainties originated in perturbation theory, we vary the scale at zero flow time with a factor $s$ as $\mu=1/\sqrt{sr^2}$ with center value $s=1$ and a variation from $s=1/2$ to $s=2$. The s-scale error covers the resulting variation of $\Lambda_0$. The scaling behavior at finite flow time, parametrized by $b$, introduces another perturbative uncertainty. We choose $b=0$ as the mean value to match the previous studies at zero flow time and vary $b$ from $b=-1/2$ to $b=1$. Finally, we obtain for F1l and F3l
\begin{linenomath}
    \begin{align}
        \sqrt{8t_0}\Lambda_0^\mathrm{F1l} &= 0.814(6)^\mathrm{lattice}(_{-71}^{+120})^\mathrm{s-scale}(_{-16}^{+20})^\mathrm{b-scale} = 0.814^{+122}_{-73} \\
        \sqrt{8t_0}\Lambda_0^\mathrm{F3lLus} &= 0.629(4)^\mathrm{lattice}(^{+18}_{-25})^\mathrm{s-scale}(^{+13}_{-7})^\mathrm{b-scale} = 0.629^{+22}_{-26}
    \end{align}
\end{linenomath}
where $()^\mathrm{lattice}$ contains the error from the lattice, consisting of the statistical error and the error from the Akaike fit window procedure. We obtain that the perturbative error decreases with increasing perturbative order. Since the $\Lambda_0$-extraction is more reliable at higher perturbative order, we report our F3lLus result as our final result.

In recent studies where $\Lambda_0$ was carried out with the gradient flow method ~\cite{Hasenfratz:2023bok, DallaBrida:2019wur}, 
the $\Lambda_0$ parameter was found to 
$\sqrt{8t_0}\Lambda=0.622(10)$ and $\sqrt{8t_0}\Lambda=0.6227(98)$, respectively, where only in the latter study a direct determination in the gradient flow scale $\sqrt{8t_0}$ was conducted, while for the first one a final conversion from $r_0$ to $\sqrt{8t_0}$ was performed. Our results are in agreement with the previous literature results within the errors.

\section{Conclusion}
We conclude that gradient flow renormalizes operators with field insertions and improves the signal-to-noise ratio. This allows us to perform reliable continuum limits at finite flow time. Furthermore, in this way, the continuum limit of the direct force measurement on the lattice can be carried out and used to extract $\Lambda_0$. This can be organized by either performing the constant zero flow time limit of the force first, followed by the fit of the perturbative expression to the $\tau_F=0$ results; this $\tau_F\rightarrow 0$-limit also works in the non-perturbative large $r$ regime. Alternatively, we can fit the perturbative expressions to the data at finite flow time, where we model the flow time behavior with an arbitrary order at zero flow time and the finite flow time effects with the 1-loop expression. As the final result, we report $\sqrt{8t_0}\Lambda_0=0.629^{+23}_{-34}$ where the final error includes the statistical and the perturbative uncertainties.

\section*{Acknowledge}
In analysis, the numerical running of $\alpha_\mathrm{s}$ was performed using the \texttt{RunDec} package~\cite{Chetyrkin:2000yt,Schmidt:2012az,Herren:2017osy}. 
The simulations were carried out on the computing facilities of the Computational Center for Particle and Astrophysics (C2PAP) in the project 'Calculation of finite T QCD correlators' (pr83pu) and of the SuperMUC cluster at the Leibniz-Rechenzentrum (LRZ) in the project 'The role of the charm-quark for the QCD coupling constant' (pn56bo).
J. M.-S. acknowledges support by the Munich Data Science Institute (MDSI) at the Technical University of Munich (TUM) via the Linde/MDSI Doctoral Fellowship program.
This research was funded by the Deutsche Forschungsgemeinschaft (DFG, German Research Foundation) cluster of excellence “ORIGINS” (\href{www.origins-cluster.de}{www.origins-cluster.de}) under Germany’s Excellence Strategy EXC-2094-390783311. 

\let\oldthebibliography\thebibliography
\let\endoldthebibliography\endthebibliography
\renewenvironment{thebibliography}[1]{
  \begin{oldthebibliography}{#1}
    \setlength{\itemsep}{0em}
    \setlength{\parskip}{0em}
}
{
  \end{oldthebibliography}
}
\bibliographystyle{JHEP}
\bibliography{force.bib}

\end{document}